# Preliminary demonstration of flexible dual-energy X-ray phase-contrast imaging

**Shenghao Wang,[1,3] Guibin Zan,[2,3] Qiuping Wang,[2] Shijie Liu[1,*]**

[1]Chinese Academy of Sciences, Shanghai Institute of Optics and Fine Mechanics, Precision Optical manufacturing and Testing Center, Hui wang east road, no. 899, Shanghai, China, 201800.
[2]University of Science and Technology of China, National Synchrotron Radiation Laboratory, He zuo hua south road, no. 42, Hefei, China, 230026.
[3]These authors contributed equally.

**Abstract**. Currently, dual-energy X-ray phase contrast imaging is usually conducted with an X-ray Talbot-Lau interferometer. However, in this system, the two adopted energy spectra have to be chosen carefully in order to match well with the phase grating. For example, the accelerating voltages of the X-ray tube are supposed to be respectively set as 40 kV and 70 kV, with other energy spectra being practically unusable for dual energy imaging. This system thus has low flexibility and maneuverability in practical applications. In this work, dual energy X-ray phase-contrast imaging is performed in a grating-based non-interferometric imaging system rather than in a Talbot-Lau interferometer. The advantage of this system is that, theoretically speaking, any two separated energy spectra can be utilized to perform dual energy X-ray phase-contrast imaging. The preliminary experimental results show that dual-energy X-ray phase contrast imaging is successfully performed when the accelerating voltages of the X-ray tube are successively set as 40 kV and 50 kV. Our work increases the flexibility and maneuverability when employing dual-energy X-ray phase-contrast imaging in medical diagnoses and nondestructive tests.

**\*** Shijie Liu, E-mail: shijieliu@siom.ac.cn

# 1 Introduction

Compared with conventional dual-energy X-ray attenuation contrast imaging,[1-4] dual-energy X-ray phase-contrast imaging has great advantages when identifying, discriminating and quantifying materials in weakly absorbing samples.[5-7]

Until now, in contrast to other techniques of performing X-ray phase-contrast imaging, the X-ray Talbot-Lau interferometer has shown great potential in medical diagnoses and nondestructive tests because it demonstrates that phase-contrast X-ray imaging can be successfully performed with a conventional, low-brilliance X-ray source.[8-13] However, generally speaking, for an X-ray Talbot-Lau interferometer, because the phase shift and the Talbot distance of the phase grating are directly related to the wavelength of the X-ray, the accelerating voltage of the X-ray tube is chosen such that the mean energy of the energy spectrum emitted from the X-ray source matches the designed energy of the phase grating.[8, 14] For example, when the designed energy of an X-ray Talbot-Lau interferometer is 27 keV, the tube voltage of the X-ray source should be set in the range of approximately 35-45 kV, meaning the system does not work efficiently when the X-ray tube is operated at a tube voltage of 60 kV.[14] Therefore, conducting dual-energy imaging with an X-ray Talbot-Lau interferometer is not straightforward.[6] Note that the high energy and the low energy chosen in the dual-energy imaging should be sufficiently different in order to obtain considerably different signals, for example, 40 kV and 70 kV.[6] In 2007, Popescu Heismann et al. introduced a dual-energy X-ray phase-contrast imaging system, which consisted of two X-ray Talbot-Lau interferometers: one interferometer was used to obtain the image of the sample at a high photon energy, while the other one worked in the low photon energy range.[15] In 2010, Kottler Revol et al. successfully conducted dual-energy X-ray phase-contrast imaging with a single X-ray Talbot-Lau interferometer.[6] Two tube voltages (40 kV and 70 kV) of the X-ray tube were carefully chosen to respectively produce phase shifts of $\pi/2$ and $\pi/4$ when the X-ray penetrated through the phase grating, and a Talbot self-image of the phase grating appeared coincidently in both cases in the plane of the analyzer grating. The phase signal of the sample could thus be retrieved by analyzing the moiré fringe. It is quite clear that in the aforementioned two configurations for dual-energy X-ray phase contrast imaging, the available energy spectra (corresponding to the X-ray tube voltage) is almost fixed, and the other energy spectra are essentially unusable. However, when dealing with a sample possessing different characteristics, it is important to choose different energy spectra for the dual-energy imaging.[16] Therefore, performing dual-energy X-ray phase-contrast imaging with an X-ray Talbot-Lau interferometer is not flexible.

In this work, dual-energy X-ray phase-contrast imaging is performed in a grating-based non-interferometric imaging system (its theory has been reported before[17]), rather than in an X-ray Talbot-Lau interferometer. The advantage of this configuration is that, theoretically speaking, any two separated energy spectra can be utilized to perform the dual-energy imaging. The preliminary experimental results will be demonstrated when the accelerating voltages of the X-ray tube are set as 40 kV and 50 kV.

## 2    Measurement theory of dual-energy X-ray phase-contrast imaging using a grating-based non-interferometric imaging system

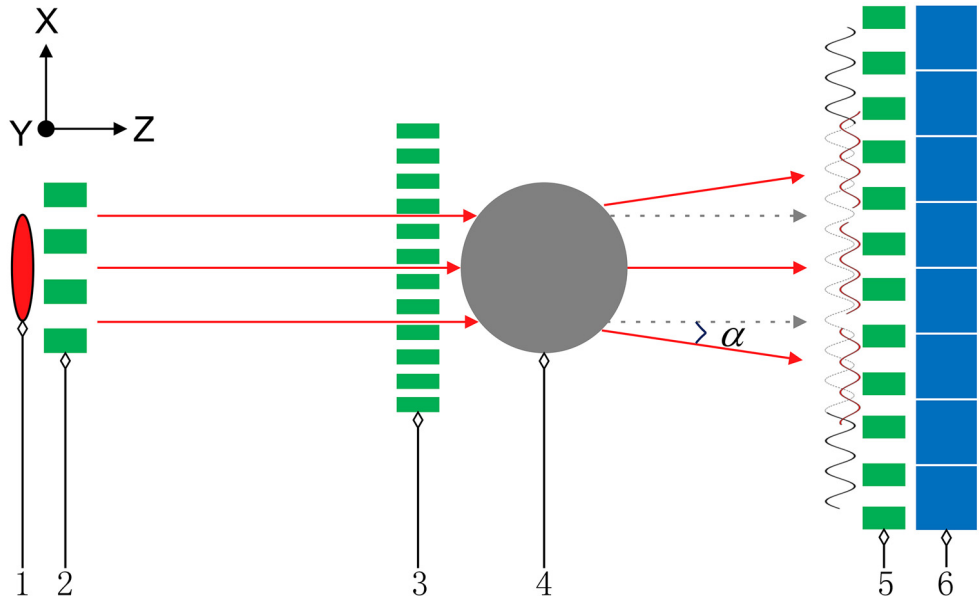

**Fig. 1** Layout of the grating-based non-interferometric X-ray phase contrast imaging system. 1: X-ray tube, 2: source grating, 3: middle grating, 4: sample, 5: analyzer grating, 6: X-ray detector.

The theory of performing X-ray phase-contrast imaging with a grating-based non-interferometric imaging system has been reported before;[17] herein, we will emphasize on introducing its differences with an X-ray Talbot-Lau interferometer and explaining how to utilize it in order to realize flexible dual-energy imaging.

Fig. 1 is the layout of the grating-based non-interferometric X-ray phase contrast imaging system. It mainly consists of an X-ray tube, a source grating, a middle grating, an analyzer grating and an X-ray detector. Its layout is almost the same as that of an X-ray Talbot-Lau interferometer;[9] the difference is that in an X-ray Talbot-Lau interferometer, the middle grating is usually a phase grating, its period is typically on the micron scale, and when it is illuminated by an X-ray beam produced by the X-ray source and the source grating, the lateral coherent length of the X-ray is larger than the period of the middle grating, so a self-image of the middle grating appears in the plane of the analyzer grating via the Talbot effect,[18] while in a grating-based non-interferometric imaging system, the middle grating is an absorption grating and its period is usually on the scale of tens of microns.[17] When it is illuminated by the X-ray beam after penetrating through the source grating, the lateral coherent length of the X-ray is smaller than the period of the middle grating, so its image is produced in the plane of the analyzer grating via direct geometric projection. When conducting X-ray phase contrast imaging with an X-ray Talbot-Lau interferometer, because the Talbot distance of the phase grating is directly related to the photon energy of the X-ray, the accelerating voltage of the X-ray tube is therefore often chosen such that the mean energy of the energy spectrum matches the Talbot distance (the distance between the middle grating and the analyzer grating) of the phase grating.[8] In contrast, in a grating-based non-interferometric imaging system, because the self-image of the middle grating is formed via a direct geometric projection, the distance between the middle grating and the analyzer grating is not related to the photon energy of the X-ray.[17] Theoretically speaking, any two separated energy spectra can be utilized to perform X-ray phase-contrast imaging. Therefore, compared with an X-ray Talbot-Lau interferometer, it is very clear that flexible dual-energy X-ray phase-contrast imaging can be performed in a grating-based non-interferometric imaging system.

## 3  Experimental setup

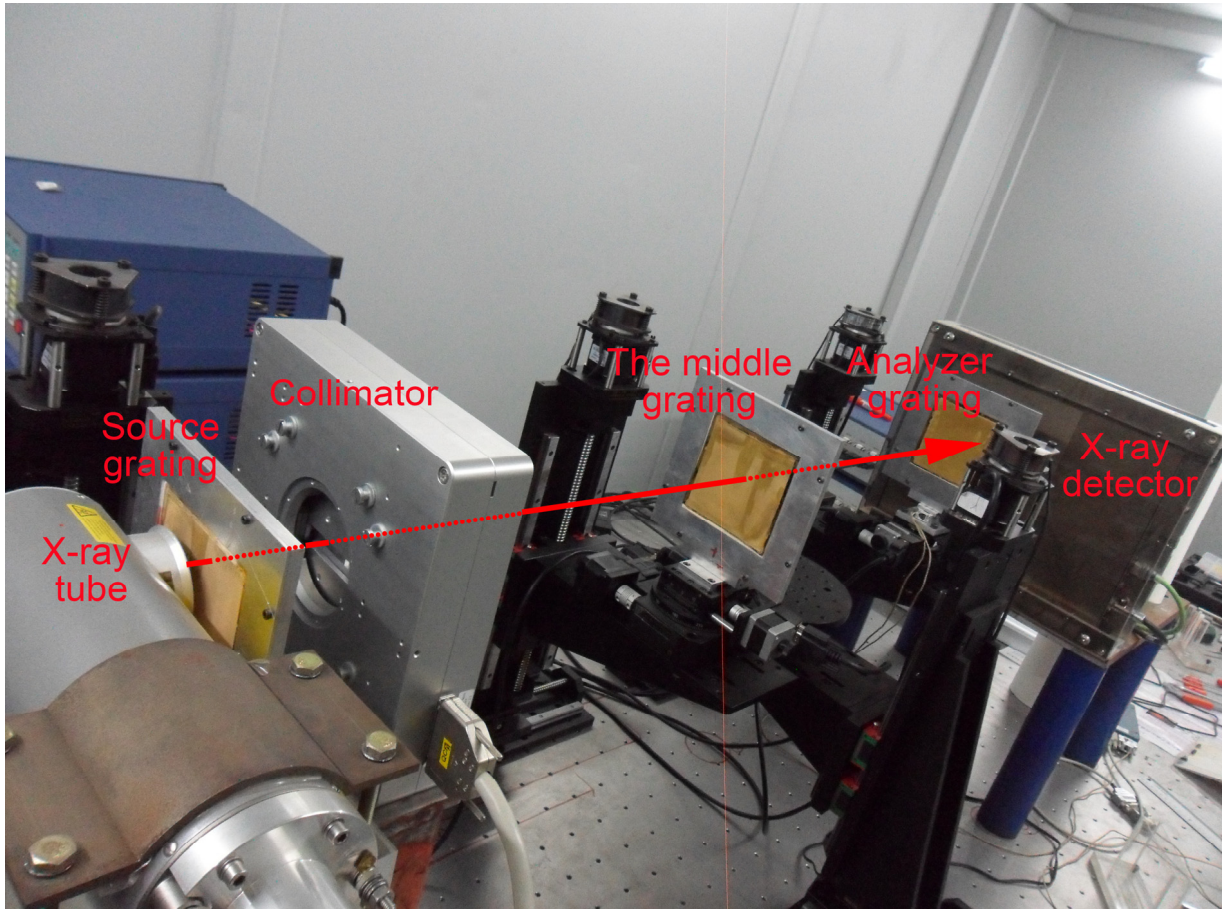

**Fig. 2** Photograph of the grating-based non-interferometric X-ray phase-contrast imaging system.

The dual-energy X-ray phase-contrast imaging was performed at the National Synchrotron Radiation Laboratory of the University of Science and Technology of China.[19] Fig. 2 is a photograph of the imaging setup. It mainly consists of an X-ray tube, an X-ray flat panel detector and three microstructured gratings, which are built on multidimensional optical stages. The X-ray tube is a cone beam source with a round focal spot (1.0 mm in diameter) on a tungsten target anode. Its operating voltage ranges from 7.5 kV to 160 kV, and the X-ray tube is cooled using a centrifugal chiller. The source grating, with a period of 100 μm and a duty circle of 0.5, is positioned approximately 10 mm behind the emission point of the X-ray source.

The middle grating, with a period of 50 μm and a duty circle of 0.5, is placed 270 mm behind the source grating along the light path. The period of the analyzer grating is 100 μm and its duty circle is 0.5, and it is positioned in contact with the flat-panel detector, where the distance between the middle grating and the analyzer grating is 270 mm. The X-ray flat panel detector has an effective receiving area of 20.48×20.48 cm$^2$ and its pixel size is 0.2×0.2 mm$^2$. The system is automatically controlled by a custom software.[20]

The samples were polymethylmethacrylate (PMMA) and polyformaldehyde (POM) cylinders, both of which had two external dimensions (diameters of 0.5 cm and 1 cm). After fine alignment of the three gratings, the X-ray tube was operated with a tube current of 22.5 mA. The data was collected by the popular phase stepping method,[8, 21, 22] in which 100 steps were adopted. For each step, 30 raw images were captured to reduce the statistical noise, and the exposure time of each image was 2 seconds. The phase stepping scan was repeated twice, before and after inserting the sample into the beam path. The first measurement was performed when the accelerating voltage of X-ray tube was 50 kV, and the second one was performed at a tube voltage of 40 kV.

## 4  Preliminary experimental result and quantitative analysis

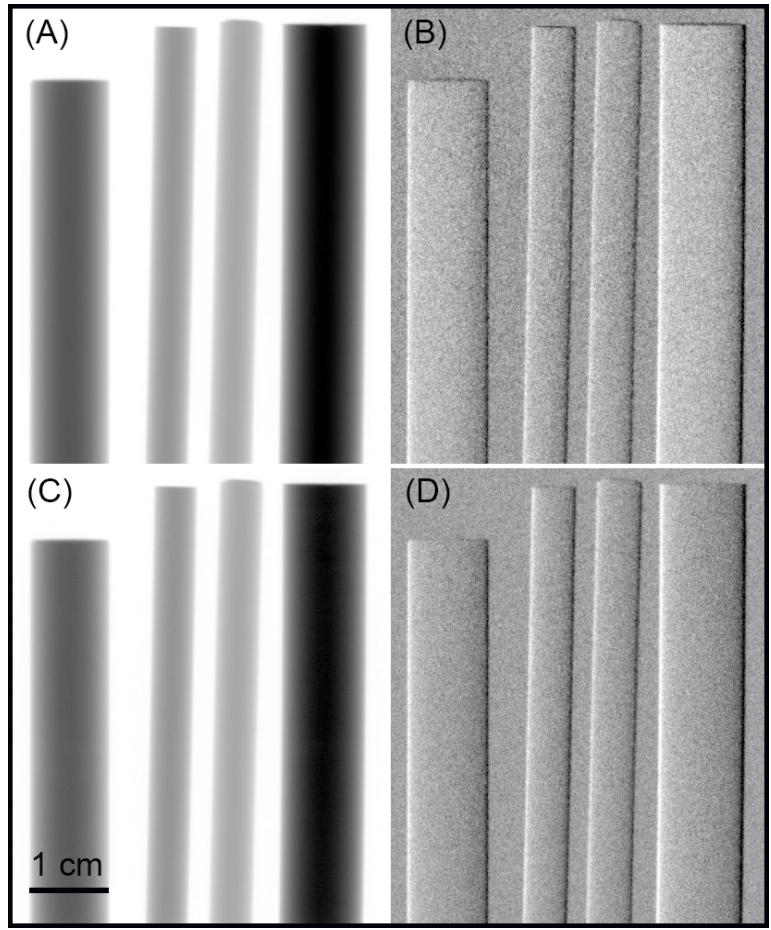

**Fig. 3** Preliminary experimental results of the dual-energy X-ray phase-contrast imaging. (A) is the absorption image and (B) is the phase contrast image, which are obtained when the X-ray tube is operated at an accelerating voltage of 50 kV. (C) and (D) are the corresponding images at a tube voltage of 40 kV. From left to right, the sample materials are PMMA, POM, PMMA and POM.

Fig. 3 demonstrates the preliminary experimental results of the dual-energy X-ray phase-contrast imaging, Fig. 3(A) is the absorption image and Fig. 3(B) is the phase-contrast image, which are retrieved by the raw phase stepping data obtained when the X-ray tube is operated at an accelerating voltage of 50 kV. Fig. 3(C) and Fig. 3(D) are the corresponding images at the accelerating voltage of 40 kV. From left to right, the sample materials are PMMA, POM, PMMA and POM.

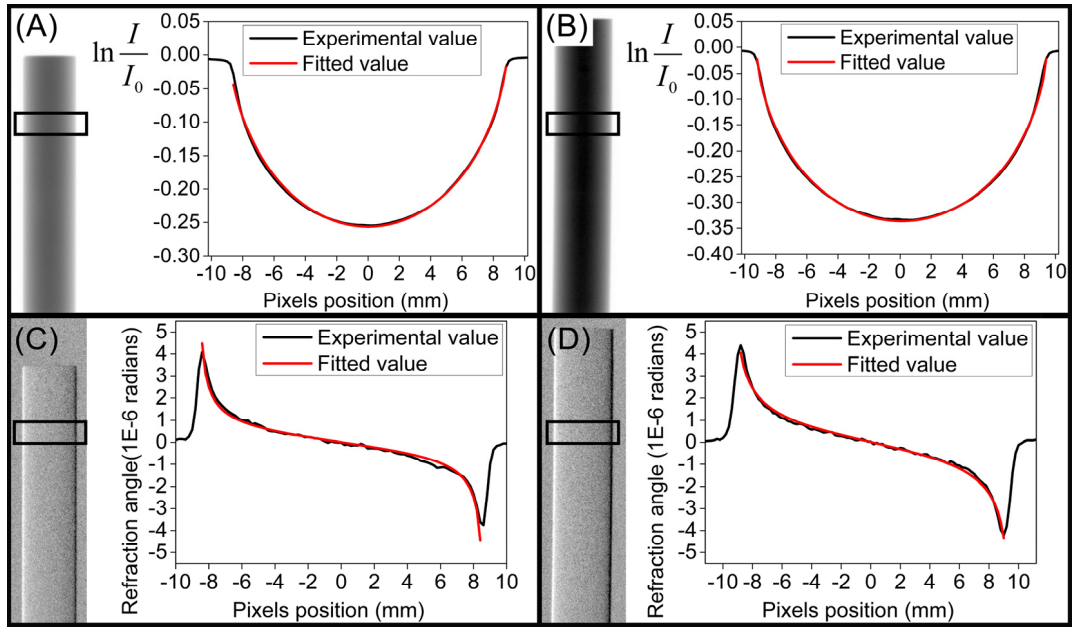

**Fig. 4** Cross-section profiles of the absorption and the refraction images when the tube voltage is 40 kV. (A) Absorption of the PMMA cylinder, (B) absorption of the POM, (C) refraction of the PMMA cylinder, and (D) refraction of the POM cylinder. The black curves are the experimental data, and the red curves are the fitted data.

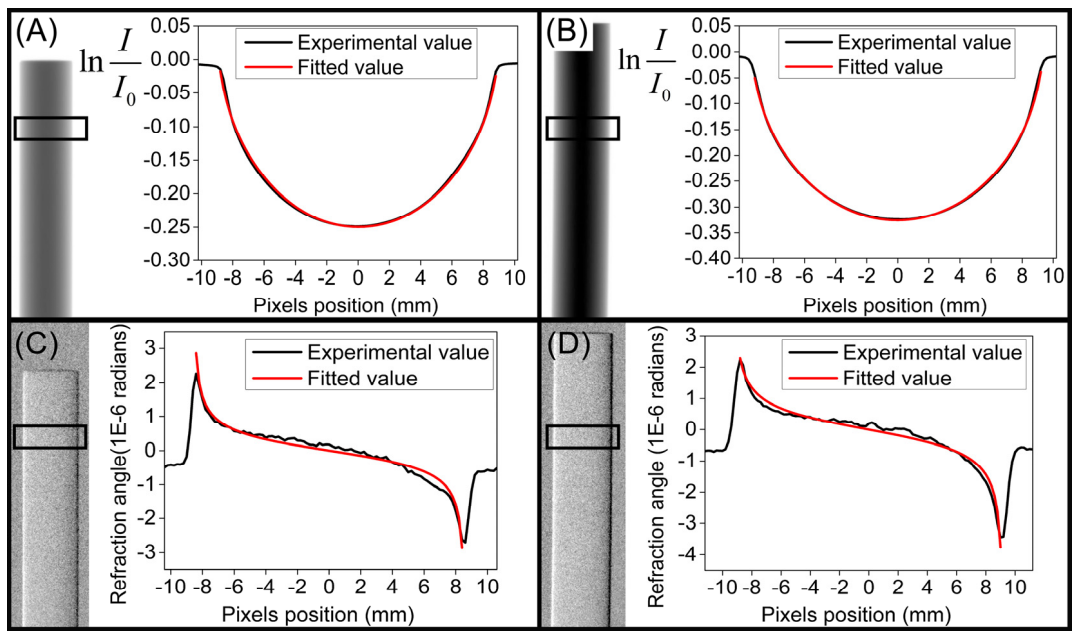

**Fig. 5** Cross-section profiles of the absorption and the refraction images when the tube voltage is 50 kV. (A) Absorption of the PMMA cylinder, (B) absorption of the POM, (C) refraction of the PMMA cylinder, and (D) refraction of the POM cylinder. The black curves are the experimental data, and the red curves are the fitted data.

The quantitative analyses of the dual energy experimental results are shown in Fig. 4 and Fig. 5. Here, a PMMA cylinder and a POM cylinder 1 cm in diameter were used for the analysis, see the left and right cylinders in Fig. 3(A). Fig. 4 shows the cross-section profiles of the absorption and the refraction images when the tube voltage of the X-ray source is 40 kV. Fig. 4(A, B, C, and D) represent the absorption of the PMMA cylinder and the POM cylinder, and the refraction of the PMMA cylinder and the POM cylinder, respectively. Fig. 5 shows the corresponding cross-section profiles when the tube voltage of the X-ray source is 50 kV, where the black curves are the experimental data, and the red curves are the fitted data.

The equation we adopted to fit the cross-section profiles in Fig. 4(A), Fig. 4(B), Fig. 5(A) and Fig. 5(B) is

$$\ln(\frac{I}{I_0}) = \frac{-4\pi\beta}{\lambda} \times 2\sqrt{R^2 - x^2}. \tag{1}$$

In Fig. 4(C), Fig. 4(D), Fig. 5(C) and Fig. 5(D), the fitting equation is

$$\theta = \frac{2\delta x}{\sqrt{R^2 - x^2}}. \tag{2}$$

In equations (1) and (2),[14, 23] $I$ and $I_0$ are the photon intensities of the X-ray beam with and without the sample, respectively, $\lambda$ is the mean wavelength of the energy spectrum (here 27 keV and 30 keV are regarded conventionally as the mean energy of the energy spectrum when the accelerating voltages of the X-ray tube are 40 kV and 50 kV, respectively), $R$ is the radius of the cylinder, $\theta$ is the measured refractive angle, $\delta$ is the real part and $\beta$ is the imaginary part of the material's refractive index, and $x$ is the variable.

Through mathematical fitting, $\delta$ and $\beta$ of the PMMA and POM can be calculated, which are listed in Table 1. We can see that preliminary dual-energy X-ray phase contrast imaging is successfully performed when the accelerating voltages of the X-ray tube are successively set as 40 kV and 50 kV.

**Table 1** $\delta$ and $\beta$ of the PMMA and POM at the two tube voltages.

| Material | Voltage of the X-ray tube | Refractive index |
|---|---|---|
| PMMA | 40 kV | $\delta=5.40\times10^{-7}$, $\beta=1.02\times10^{-10}$ |
|  | 50 kV | $\delta=3.14\times10^{-7}$, $\beta=0.87\times10^{-10}$ |
| POM | 40 kV | $\delta=5.70\times10^{-7}$, $\beta=1.25\times10^{-10}$ |
|  | 50 kV | $\delta=3.50\times10^{-7}$, $\beta=1.05\times10^{-10}$ |

## 5  Discussion

The aforementioned experimental results and quantitative analysis successfully show that flexible dual-energy X-ray phase-contrast imaging could be efficiently performed with a grating-based non-interferometric imaging system. Here we also want to point out that, when developing dual energy X-ray phase contract imaging machines in clinical diagnosis or industrial applications, compared with the currently popular X-ray Talbot-Lau interferometer, the adoption of non-interferometric imaging framework has the following other potential features apart from the flexibility of choosing the dual energy.

(a) The large area (around $40\times40$ cm$^2$ is often needed for human body detection) and high absorption thickness (about 400 μm is typically required to produce sufficient absorption of the high energy X-ray photons, for example 80 keV, note that lower energy photons usually cannot penetrate the human body and other large samples[24]) grating would be easier to fabricate, because the depth-width ratio of the absorption grating is much smaller when the absorption thickness is the same, note that manufacturing high depth-width ratio X-ray absorption grating is really not easy;[25] this feature means that developing dual energy X-ray phase contrast facility in clinical applications and in industrial non-destructive test would be more achievable.

(b) Considering that the dual energy X-ray phase contrast imaging system is developed based on gratings with periods of tens of microns, rather than periods of several microns in an X-ray Talbot-Lau interferometer, the system's requirement on the hardware structure (such as mechanical deformation, manufacturing error, assembly error, mechanical vibration and thermal expansion.) would be remarkably reduced, it should be noted that environmental factors usually cause uncontrollable shaking of the moire fringe in an X-ray Talbot-Lau interferometer.

(c) The distance among the three gratings can be flexibly adjusted, on the condition that a simple geometrical relationship is guaranteed, note that in an X-ray Talbot-Lau interferometer, for a giving phase grating, the relative positions of the main components in the system is almost fixed.

(d) The measurement range of the refractive angle is much larger, this means phase wrapping is easier to avoid when computing the sample's refractive angle. Equation (3) is often used to retrieve the refractive angle based on phase stepping in grating based X-ray phase contrast imaging.[14]

$$\alpha(x,y) = \frac{d}{2\pi l} \times \arg\left\{\frac{\sum_{k=1}^{N}\left[I_k^s(x,y) \times \exp\left(2\pi i \frac{k}{N}\right)\right]}{\sum_{k=1}^{N}\left[I_k^b(x,y) \times \exp\left(2\pi i \frac{k}{N}\right)\right]}\right\}. \qquad (3)$$

In equation (3), $\alpha(x,y)$ is the refractive angle in the pixel position of $(x,y)$, $d$ is the period of the analyzer grating, $l$ is the distance between the sample and the analyzer grating, $N$ is the number of steps during the phase stepping scan in one period of the analyzer grating, $I_k^s(x,y)$ and $I_k^b(x,y)$ are the intensity of the pixel $(x,y)$ at the $k^{th}$ step of the scan with and without sample, respectively. arg means computing the argument of a complex number, and its value range is $(-\pi,\pi]$, and it can be derived that the measurement range of the refractive angle is $(-\frac{d\pi}{2\pi l}, \frac{d\pi}{2\pi l}]$. If the desired refractive angle goes beyond this range, for example, $\frac{2d\pi}{3\pi l}$, phase wrapping would occur and complex algorithm is needed to obtain the real refractive angle;[26] thus we can say that the measurement range of the refractive angle is much larger because of the larger grating period, when compared with that in an X-ray Talbot-Lau interferometer.

Finally, we want to point out that, when choosing the high tube voltage for the dual-energy imaging in our experiment, 70 kV or 80 kV was not used, the reason is that the Au height of our absorption gratings is only 50 μm, the visibility of the moiré fringe would be unacceptable for retrieving the refractive image when our absorption gratings work in such high X-ray tube voltages, because the absorption of the Au grating would be lower when increasing the photon energy. Note that in our experiments, the visibility of the moiré fringe was 16% and 6% when the X-ray tube was operated respectively at 40 kV and 50 kV (this fact can also explain why the deviations between the experiment value and the fitted one in Fig. 5(C) and Fig. 5(D) are greater than those in Fig. 4(C) and Fig. 4(D), because higher fringe visibility usually means lower noise in the obtained phase image.[27, 28]), and if a tube voltage of 70 kV or 80 kV is applied, a lower visibility would be generated.

# 6 Conclusion

In conclusion, dual-energy X-ray phase-contrast imaging was performed with a grating-based non-interferometric imaging system. The advantage of this configuration is that, theoretically speaking, any two separated energy spectra can be utilized to perform dual-energy imaging. Our work increases the flexibility and maneuverability when employing dual-energy X-ray phase-contrast imaging in medical diagnoses and nondestructive tests.

**Acknowledgments**

This research was supported by the National Natural Science Foundation of China (Nos. 61705246, 11602280 and U1831211). The authors thank the editor (Prof. Wayne R. McKinney) and the anonymous reviewers for their constructive comments, which have improved this manuscript.

**References**

1. D. S. Maier, J. Schock, and F. Pfeiffer, "Dual-energy micro-CT with a dual-layer, dual-color, single-crystal scintillator," *Opt. Express* **25**(6), 6924-6935 (2017) [https://doi.org/10.1364/OE.25.006924].

2.  M. J. Menten et al., "Using dual-energy x-ray imaging to enhance automated lung tumor tracking during real-time adaptive radiotherapy," *Med. Phys.* **42**(12), 6987-6998 (2015) [https://doi.org/10.1118/1.4935431].

3.  M. Patino et al., "Material Separation Using Dual-Energy CT: Current and Emerging Applications," *Radiographics* **36**(4), 1087-1105 (2016) [https://doi.org/10.1148/rg.2016150220].

4.  S. Abbasi, M. Mohammadzadeh, and M. Zamzamian, "A novel dual high-energy X-ray imaging method for materials discrimination," *Nucl. Instrum. Meth. Phys. Res. A* **930**(21), 82-86 (2019) [https://doi.org/10.1016/j.nima.2019.03.064].

5.  E. Braig et al., "Direct quantitative material decomposition employing grating-based X-ray phase-contrast CT," *Sci. Rep.* **8**,16394 (2018) [https://doi.org/10.1038/s41598-018-34809-6].

6.  C. Kottler et al., "Dual energy phase contrast x-ray imaging with Talbot-Lau interferometer," *J. Appl. Phys.* **108**(11), 114906 (2010) [https://doi.org/10.1063/1.3512871].

7.  H. J. Han et al., "Preliminary research on dual-energy X-ray phase-contrast imaging," *Chin. Phys. C* **40**(4), 048201 (2016) [https://doi.org/10.1088/1674-1137/40/4/048201].

8.  T. Weitkamp et al., "X-ray phase imaging with a grating interferometer," *Opt. Express* **13**(16), 6296-6304 (2005) [https://doi.org/10.1364/OPEX.13.006296].

9.  F. Pfeiffer et al., "Phase retrieval and differential phase-contrast imaging with low-brilliance X-ray sources," *Nat. Phys.* **2**(4), 258-261 (2006) [doi:10.1038/nphys265].

10. H. X. Miao et al., "Motionless phase stepping in X-ray phase contrast imaging with a compact source," *P. Natl. Acad. Sci. USA* **110**(48), 19268-19272 (2013) [https://doi.org/10.1073/pnas.1311053110].

11. M. Hoshino, K. Uesugi, and N. Yagi, "4D x-ray phase contrast tomography for repeatable motion of biological samples," *Rev. Sci. Instrum.* **87**(9), 093705 (2016) [https://doi.org/10.1063/1.4962405].

12. S. Bachche et al., "Laboratory-based X-ray phase-imaging scanner using Talbot-Lau interferometer for non-destructive testing," *Sci. Rep.* **7**, 6711 (2017) [DOI:10.1038/s41598-017-07032-y].

13. T. J. Schroter et al., "Large field-of-view tiled grating structures for X-ray phase-contrast imaging," *Rev. Sci. Instrum.* **88**(2), 015104 (2017) [https://doi.org/10.1063/1.4973632].

14. S. H. Wang et al., "Experimental research on the feature of an x-ray Talbot-Lau interferometer versus tube accelerating voltage," *Chin. Phys. B* **24**(6), 068703 (2015) [https://doi.org/10.1088/1674-1056/24/6/068703].

15. S. Popescu, B. Heismann, and E. Hempel, "Method for producing projective and tomographic phase contrast images with the aid of an x-ray system," *US Patent* 0183560 A0183561 (2007).

16. L. Grodzins, "Optimum Energies for X-Ray Transmission Tomography of Small Samples - Applications of Synchrotron Radiation to Computerized-Tomography .1.," *Nucl. Instrum. Meth.* **206**(3), 541-545 (1983) [https://doi.org/10.1016/0167-5087(83)90393-9].

17. Z. F. Huang et al., "Alternative method for differential phase-contrast imaging with weakly coherent hard x rays," *Phys. Rev. A* **79**(1), 013815 (2009) [https://doi.org/10.1103/PhysRevA.79.013815].

18. M. V. Berry, and S. Klein, "Integer, fractional and fractal Talbot effects," *J. Mod. Opt.* **43**(10), 2139-2164 (1996) [DOI: 10.1080/09500349608232876].

19. F. Wali et al., "Low-dose and fast grating-based x-ray phase-contrast imaging," *Opt. Eng.* **56**(9), 094110 (2017) [https://doi.org/10.1117/1.OE.56.9.094110].

20. S. H. Wang et al., "A user-friendly LabVIEW software platform for grating based X-ray phase-contrast imaging," *J. X-Ray Sci. Technol.* **23**(2), 189-199 (2015) [DOI: 10.3233/XST-150480].

21. A. Momose et al., "Demonstration of X-Ray Talbot interferometry," *Jpn. J. Appl. Phys.* **42**(7b), L866-L868 (2003) [https://doi.org/10.1143/JJAP.42.L866].

22. J. H. Bruning et al., "Digital Wavefront Measuring Interferometer for Testing Optical Surfaces and Lenses," *Appl. Optics* **13**(11), 2693-2703 (1974) [https://doi.org/10.1364/AO.13.002693].

23. A. Momose et al., "Recent advances in X-ray phase imaging," *Jpn. J. Appl. Phys.* **44**(9A), 6355-6367 (2005) [https://doi.org/10.1143/JJAP.44.6355].

24. M. Willner et al., "Quantitative X-ray phase-contrast computed tomography at 82 keV," *Opt. Express* **21**(4), 4155-4166 (2013) [https://doi.org/10.1364/OE.21.004155].

25. T. Thüring et al., "X-ray phase-contrast imaging at 100 keV on a conventional source," *Sci. Rep.* **4**, 5198 (2014) [https://doi.org/10.1038/srep05198].

26. F. M. Epple et al., "Unwrapping differential x-ray phase-contrast images through phase estimation from multiple energy data," *Opt. Express* **21**(24), 29101-29108 (2013) [https://doi.org/10.1364/OE.21.029101].

27. V. Revol et al., "Noise analysis of grating-based x-ray differential phase contrast imaging," *Rev. Sci. Instrum.* **81**(7), 073709 (2010) [https://doi.org/10.1063/1.3465334].

28. K. J. Engel et al., "Contrast-to-noiseinX-raydifferentialphasecontrastimaging," *Nucl. Instrum. Meth. Phys. Res. A* **648**(S1), S202-S207 (2011) [https://doi.org/10.1016/j.nima.2010.11.169].

**Shenghao Wang** is an associate professor at the Shanghai Institute of Optics and Fine Mechanics, Chinese Academy of Sciences. He received his BS and Ph.D degrees from the University of Science and Technology of China in 2010 and 2015, respectively. His research topics involve new methods and innovative technology in the measurement of X-ray phase, X-ray absorption spectrum, UV-VIS–NIR transmittance and reflectance spectrum, spectral diffraction efficiencies of grating, and other optical parameters.